\documentclass{article}
\pdfoutput=1

\usepackage{lscape}
\usepackage{amsfonts, amssymb, amsmath, latexsym, amsthm}
\usepackage{url}
\usepackage[
	backend=biber, natbib,
	style=numeric]{biblatex}
\usepackage{float}
\usepackage{graphicx} %Graphics/figures
\usepackage{rotating} 
\usepackage{booktabs,threeparttable, multirow} 
\usepackage{tikz}
\usetikzlibrary{calc,arrows,positioning,shapes,shapes.gates.logic.US,trees, intersections}
\usepackage[T1]{fontenc} 
\usepackage{placeins}
\usepackage{hyperref}
\usepackage[hypcap=true,width=\textwidth]{subcaption}
\usepackage[margin=1in]{geometry}
\usepackage{setspace}
\title{Identifying latent classes with ordered categorical indicators}
\date{}

\author{
  R. Noah Padgett\\
  Department of Educational Psychology\\
  Baylor University\\
  Waco, TX 76706 \\
  {\tt noah\_padgett1@baylor.edu}
   \and
  Rebecca J. Tipton \\
  Department of Educational Psychology\\
  Baylor University\\
  Waco, TX 76706
}

\begin{document}
\maketitle

\begin{abstract}
A Monte Carlo simulation was used to determine which assumptions for ordered categorical data, continuity vs. discrete categories, most frequently identifies the underlying factor structure when a response variable has five ordered categories. 
The impact of infrequently endorsed response categories was also examined, a condition that has not been fully explored.
The typical method for overcoming infrequently endorsed categories in applied research is to collapse response options with adjacent categories resulting in less response categories that are endorsed more frequently, but this approach may not necessarily provide useful information.
Response category endorsement issues have been studied in Item Response Theory, but this issue has not been addressed in classification analyses nor has fit measure performance been examined under these conditions.
We found that the performance of commonly used fit statistics to identify the true number of latent class depends on the whether continuity is assumed, sample size, and convergence.
Fit statistics performed best when the five response options are assumed to be categorical. 
However, in situations with lower sample sizes and when convergence is an issue, assuming continuity and using the adjusted Lo-Mendell-Rubin likelihood ratio test may be useful.
\end{abstract}
%% Keywords
%\keywords{Finite mixture models, Fit, response frequency, polytomous items, aLMR}
\vspace*{.5cm}
Classification is an integral part of modern society. The presence of classification systems is ubiquitous; ranging from how groceries are organized in a store to grouping children by ability levels in a classroom. Classification, or clustering, of individuals in a population is oftentimes required so treatments or programs can be implemented for identified subgroups. Classification analyses aim to bring a more objective stance into such groupings. The utility of interventions is dependent on accurate group identification through appropriate use of classification analysis, thus choosing the appropriate models and discarding those that are not a good fit or representation of the sample is important \citep{Mosteller1977, Breinman2001}. Many procedures and frameworks exist to assess group belonging and how well a generated model, or theoretical mathematical representation of relationships in a population, fits the observed sample. One such framework is mixture modeling. Mixture modeling works under the premise that observed data are from a heterogeneous population and the modeling framework aims to identify groups of cases that are similar based on observed characteristics into more homogeneous subgroups. The resulting homogeneous populations are the {\it mixtures} that comprise the observed data \citep{McLachlan2000}. 

Two such analyses that aim to find homogeneous subgroups from data that are assumed to be from a heterogeneous population are latent class analysis (LCA) and latent profile analysis (LPA). These two analyses have the same goal; to classify cases (individuals) into a small number of groups. Although, each analysis makes different assumptions about the observed data, LCA assumes the indicators are categorical while LPA assumes the indicators are continuous. Prior methodological investigations have examined the effect of nonnormal indicators have on model selection and particularly on the deciding the number of latent classes \citep{McLachlan2000, Morgan2016}. 

A type of nonnormally distributed indicator is when a response category is infrequently selected; a condition that has not been fully explored. Consider an item where all the respondents answer a question such that no missing data exist for the item, but the responses could still likely be skewed and one of the categories could be unendorsed or infrequently endorsed by chance or by nature of the sample. For example, the resulting frequency of item category responses would contain a category without sufficient endorsement in the sample (i.e. sample size = 500; SD = 25, D = 0, N = 50, D = 200, SD = 225). The typical method for overcoming this occurrence in applied research is to collapse the response options with an adjacent category, resulting in a more parsimonious solution but not necessarily one that is as representative or useful. Collapsing categories is a simple solution to help with estimation of parameters to make the number of categories one less. Infrequency of response category endorsement is discussed in Item Response Theory, where the latent variable of interest is continuous. ``If a response category has zero or few responses, a program to estimate parameters will not be able to estimate [graded response model parameters]'' \citep{Embretson2000}. Although, the issue has not been fully explored in classification analyses nor has fit measure performance been examined under these conditions. 

In this study, we investigate what happens when ordered response categories are assumed continuous and mixture modeling is employed. The assumption of continuity is expected to aid in model selection when a response category is unendorsed or infrequently endorsed. Furthermore, we investigate the performance of fit statistics used in model selection under these conditions. The remainder of our discussion structures as follows. We discuss the LCA/LPA model considered in more detail, followed by an brief introduction to the statistical fit measures for model selection included. We then describe the Monte Carlo simulation study conditions. The results are in the third section followed by a discussion that includes recommendations for use of statistical fit indices for model selection. 

\subsection*{LCA and LPA}
Classification analyses such as Latent Class Analysis (LCA) and Latent Profile Analysis (LPA) are two types of mixture models used for classification. These models are receiving growing attention in methodological research due to the expanding application of such models \citep{Williams2016, Morgan2015, Nylund2007, Soromenho1994}. The goal of LCA and LPA alike is to classify similar objects into one of K groups or classes of unknown form and frequency, with the form of the group referring to cluster-specific centroids, variances, and covariance, and the frequency referring to the number of underlying groups present.

The model is briefly introduction below; however a more in-depth discussion of the LCA/LPA models can be found in \cite{McLachlan2000}.
Let \( \mathbf{y}_i \) represent a vector of responses of individual $i$ on the $j^{th}$ item, where \( i = 1,2, ... , N\) for the number of individuals sampled and \( j = 1,...,K\) for $K$ being the number of latent classes specified. 
LCA is a probabilistic model for each unique response pattern observed in a sample and is defined as
% LCA Model:
\begin{equation}
	P(\mathbf{y}_i) = \sum_{k=1}^{K}{P( X=k)}\ \prod_{j=1}^{J}{P(y_{ij} \ | X=k)}
\end{equation}
where \( P(\mathbf{y}_i)\) is the probability of response pattern for individual $i$, \( P( X=k) \) is the probability of membership in class $k$ also known as class prevalence which is the proportional size of class $k$, and \( P(y_{ij} | X=k)\) is the probability of the response of individual $i$ on the $j^{th}$ item conditional on class membership. The difference between LCA and LPA is in the form of the item response probability. In LCA, each class has specific item response/endorsement probabilities that differential the classes, whereas in LPA each class has specific response means and variances.

\subsection*{Model Selection Assumptions}
Models, by definition, are only approximations to unknown reality or truth. George Box made the famous statement, ``All models are wrong but some are useful.'' Models that perfectly reflect reality do not exist, but selecting models that best represent relationships in the population can still be useful in decision-making. Three general guiding principles that should be considered during model selection are (a) parsimony, (b) multiplicity or alternative model hypotheses, and (c) strength of evidence. Inference under models with too few parameters can be biased, but models having too many parameters results in poor precision and spurious conclusions. A balance must exist between under- and over-specification during the model selection process. As such, several working hypotheses or alternate models must be considered as potential representations of the true population based on theory and previous research, and assessed accordingly, but the number should be kept small and should consider scientific context. Judgment on the part of the researcher is critical during the hypothesis testing stage, and considering multiple sources and strength of evidence collected in concert with previous research findings is important in model selection.

\subsection*{Fit Indices for Model Selection}
Once models have been estimated, the researcher is tasked with selecting the optimal fitting model. The model with the correct number of latent classes is selected using a mix of statistical evidence and guiding theory. There are no established guidelines to determine adequate fit of a theoretical model to data, but ensuring that parameter estimates are within reasonable range and standard errors for the estimates are not too large is an important component of considering model fit \citep{Marsh1995}. The fit indices included in a study are suggested to be chosen based upon the study being conducted, but cannot be evaluated independently of one another \citep{Tanaka1993}. In this article, we examine the methods for determining fit using two major procedures of model selection statistical using information criteria and likelihood ratio based tests.

\subsubsection*{Information Criteria Procedures}

Measures that use the information criteria are of particular importance because they provide a framework for comparing models with differing numbers of parameters and different class enumerations. In general, the model that has the lowest information criteria in terms of absolute value, is the model that best approximates the relationships observed in the population, and models that do not fit better than the baseline model can be dismissed. For this study, we have chosen to assess Akaike's Information Criterion (AIC), corrected AIC (AICc), Bayesian Information Criterion (BIC), and sample size-adjusted Bayesian Information Criterion (ssBIC). These criteria are sensitive to parameters like sample size and parsimony, and model selection should be informed by biases innate in the criteria selected. To date, there is not common acceptance of the best criteria for determining the number of classes in mixture modeling, despite various suggestions.

To see how well these models fit collected data, the log-likelihood value is typically used. Due to the likelihood function taking on small values and the resulting log likelihood being a negative value, a value close to 0 indicates optimal fit. A log-likelihood closer to 0 is the same as the likelihood function approaching 1, which would indicated that the model predicts these data well. Prior examination of correct class enumeration based on the log-likelihood values, however, has shown poor results \citep{Nylund2007, Morgan2015}.

The Akaike Information Criterion (AIC) is a goodness-of-fit measure that reflects the extent to which the observed covariance matrix varies from the model or predicted covariance matrix \citep{Akaike1974, Akaike1987}, with a lower AIC value for two competing models indicating better model fit (Kaplan, 2000; Takane \& Bozdogan, 1987). 
The AIC is defined as 
\begin{equation}
AIC = -2 \log L + 2p
\end{equation}
Where \(L\) is value from the likelihood function and \(p\) is the number of free model parameters. The AIC value is penalized for complexity of the proposed model, meaning use of the AIC is an attempt to minimize the overall error caused by added parameters. 
The corrected AIC (AICc) is biased corrected version of the AIC under small sample sizes \citep{Sugiura1978, Burnham2004}. The AICc is considered to be more stringent than the AIC, having a greater penalty for models having larger numbers of parameters.
\begin{equation}
AICc = -2 \log L + 2p + \frac{2p(p+1)}{n-p-1}
\end{equation}
The Bayesian Information Criteria (BIC) is similar to AIC in that it is also a measure of goodness-of-fit as well as the extent to which the observed covariance matrix varies from the predicted covariance matrix \citep{Schwarz1978}. The BIC is defined as
\begin{equation}
BIC = -2\log L+p \log (n)
\end{equation}
A lower BIC value for two competing models indicates a better model fit; the penalty of BIC increases with sample size.
To accommodate for this penalty, the sample size-adjusted Bayesian Information Criterion (ssBIC) replaces n in the above equation with \(n^{*} = (n+2)/24\)
\begin{equation}
ssBIC = -2 \log L + p \log [(n^* +2)/24]
\end{equation}
For each of the information criteria, the optimal fitting model is chosen based on which of the compared model have the lowest value. 
\subsubsection*{Likelihood Ratio Tests}

Other measures of fit include the Lo-Mendell-Rubin likelihood ratio tests. We focus our attention on the adjusted Lo-Mendell-Rubin Likelihood Ratio Test (aLMR)  \citep{Lo2001, Vuong1989}. The likelihood ratio test is able to compare models that differ in the number of classes by indicating that the model with K-1 classes should be rejected in favor of the model with K classes (Lo, Mendell, \& Rubin, 2001; Muthen, 2004). The aLMR is sample-size dependent, meaning that a larger sample size inflates the test statistics \citep{Lo2001}. 

Bootstrapping procedures can also be used to supplement the likelihood ratio test family by generating and using empirical distributions of likelihoods \citep{McLachlan2000}. Comparison of models using a likelihood-based technique can be done by a parametric bootstrapping method where bootstrap samples are used to estimate differences in the distribution and a p-value allows for comparison of the K-1 and K class models. The bootstrap likelihood ratio test (BLRT) can then be used to determine if two competing models are significantly different, as described in detail by \cite{Nylund2007, Asparouhov2012}. It has been suggested that the BLRT may be more appropriate than LMR statistics when comparing LCA models due to BLRT's accuracy and tendency to produce nonsignificant findings for models with increased K, whereas LMR findings tend to show significance, then insignificance, and then significance again as k increases.  

Both the aLMR and BLRT procedures will be generated in this study to provide a value for assessment of whether the proposed K-1 class model should be rejected in favor of the K class model. The K-1 class is selected as optimal if the aLMR p-value for the test of K-1 to K is found to be greater than .05, meaning that the K class does not provide a better fit than the K-1 class. 

\section*{Methods}
\subsection*{Population Model Structure}
Conditions were selected that mirror empirical research situations as closely as possible. The population structures for this study were replicated from \cite{Morgan2015}, where Morgan reviewed educational databases for studies using LCA or similar mixture modeling techniques. The three factors varied were sample size, class prevalence, and category response distribution. The included sample sizes are 500, 1000, and 1500.
\subsubsection*{Class Prevalence} 
The class prevalence or class size is representative of various sizes of the latent populations. The underlying populations may be approximately equal in size or a class may be more prevalent than the others. The existence of a rare class might also be speculated based on guiding theory; for example, in psychiatric disorders the class exhibiting the disorder is likely a small portion of the general population. Varying class prevalence changes the probability of accurately defining a class, and classes that are rare are harder to define or statistically justify.
Three underlying classes are assumed in our study and the class prevalences considered are 0.45-0.40-0.15, 0.59-0.26-0.15, and 0.89-0.08-0.03.
\subsubsection*{Category Response Distribution} 
The main scope of our investigation is to assess the category response distribution when obtained data are ordered. We assumed the collection of data from ten items each have five ordered categories. Response probability is the primary method for classifying people based on like responses, so if a category is unendorsed or infrequently endorsed, classification becomes more difficult because less variability exists in response patterns. In our study, three different conditions are compared. The first condition has a distribution that contains an unendorsed response in the second the lowest category. The second condition contains a category that is infrequently endorsed with only 1\% of subjects in each class endorsing that category. The last condition is a control condition where each pf the categories are endorsed in a normal distribution. The distribution of class category response probabilities for the indicators are given in Table \ref{tb:respdist}.
\begin{table}[ht]
\centering
\begin{threeparttable}
\caption{Distribution of class response probabilities per condition}
\label{tb:respdist}
	\begin{tabular}{l*{6}{c}}\toprule %18
	Response						&				& \multicolumn{5}{c}{Response Categories}	 \\  \cmidrule(r){3-7}
	Frequency\tnote{1}			& 		Class	&	SD 	&	D		&	N	  	&	A		&	SA	  \\ \midrule
	Unendorsed	&			1	& .05		& .00	& .05  		& .10	& .80  \\
	 	&			2	& .05		& .00	& .75		&	.15	&	.05 \\
		&			3	& .90  		& .00	& .05		& .025	& .025 \\
	Infrequent	&			1	& .049		& .01	& .05		& .10	& .80\\
		&			2	& .05		& .01	& .75		& .149	&	.05	\\
		&			3	& .90 		& .01 	& .049		& .025	& .025\\
	Control	&			1	&	.025		&.025	& .05	& .20		& .70\\
	 	&			2	&	.05		& .10	& .70	& .10		& .05\\
		&			3	&	.70		& .20	& .05	& .025		& .025\\\bottomrule
	\end{tabular} 
	\begin{tablenotes}[para, flushleft]
    {\small
        \textit{Note.}
		${}^{1}$condition designating how endorsed each item response is across conditions; 1) a response category is not endorsed, 2) infrequently endorsed response category, and 3) control with all response options selected. 
	}
	\end{tablenotes}
	\end{threeparttable}
\end{table}

\subsection*{Summarizing Results}
Mirroring real-life applied research problems when trying to identify the unknown number of classes, we fit one through five class solutions for each type of analysis. Model estimation was conducted in M{\it plus} 8.1  using the MIXTURE option \citep{Muthen2017}. The optimal number of classes for each condition was based on the decision rule for each fit measure. For each fit measure, if a three class solution is found to be optimal then the solution is flagged as correct with a one while if another other class enumeration is found to be optimal then the solution is flagged as incorrect with zero. We summarize the proportion of replications that each fit indice correctly identified the number of latent classes. Using logistic regression, we analyzed which condition(s) and/or interaction of conditions had the largest effect on model selection. The logistic regression model tested uses the three factors described above and the analysis type (LCA vs. LPA) as indicators. \cite{Hancock2013} suggest the use of inferential statistics in simulation to aid in result interpretation. All analyses after model estimation was completed using R \citep{r2017}.

\section*{Results}
The entirety of the results of this simulation study are available online \citep{Padgett2019}.
A total of 27,000 data sets were successfully generated, 27(3x3x3) conditions by 1,000 replications. For each data set one- through five- class solutions were tested using categorical and continuous indicator assumptions, i.e. LCA and LPA. Testing these models for each of 1000 replicates in each of 27 conditions yields 270,000 models were estimated. Convergence was found to be problematic when assuming categorical indicators, especially in the infrequently endorsed category conditions. Models that did not converge were flagged and the replication was deemed unusable, an approach taken in similar simulation studies, and nonconvergence is typically due to an ill-defined likelihood function and/or an insufficient number of random starts \citep{Nylund2007, Morgan2015}. 
The convergence rates for each condition are provided in Table \ref{tb:converge}.
\begin{table}[ht]
\centering
\begin{threeparttable}
\caption{Convergence rate across replications}
\label{tb:converge}
\begin{tabular}{l*{7}{c}}
  \toprule
  Response	 &	Class	&	\multicolumn{2}{c}{N=500} &	\multicolumn{2}{c}{N=1000} &	\multicolumn{2}{c}{N=1500}	 \\  \cmidrule(lr){3-4} \cmidrule(lr){5-6} \cmidrule(lr){7-8}
  Frequency  & Prevalence & LCA & LPA & LCA & LPA &  LCA & LPA \\ 
  \midrule
  Unendorsed & .45-.40-.15 & 1.000 & 1.000 & .990 & 1.000 & .967 & 1.000 \\ 
   & .59-.26-.15 & .998 & 1.000 & .994 & 1.000 & .970 & 1.000 \\
   & .89-.08-.03 & .999 & 1.000 & .979 & 1.000 & .898 & 1.000 \\ 
  Infrequent & .45-.40-.15 & .996 & 1.000 & .935 & 1.000 & .826 & 1.000 \\ 
   & .59-.26-.15 & .998 & 1.000 & .967 & 1.000 & .875 & 1.000 \\ 
   & .89-.08-.03 & .996 & 1.000 & .938 & 1.000 & .736 & 1.000 \\
  Control & .45-.40-.15 & .999 & 1.000 & .981 & 1.000 & .931 & 1.000 \\
   & .59-.26-.15 & .999 & 1.000 & .985 & 1.000 & .949 & 1.000 \\ 
   & .89-.08-.03 & .999 & 1.000 & .958 & 1.000 & .801 & 1.000 \\ 
   \bottomrule
\end{tabular}
\end{threeparttable}
\end{table}

\subsection*{Information Criteria}
In this study, four different information criteria were examined for model selection, the Akaike information criterion (AIC), corrected Akaike information criterion (AICc), Bayesian information criterion (BIC), and sample size-adjusted (ssBIC). The proportion of valid replications that each index correctly identified the number of latent classes are reported in Table \ref{tb:results}. The results from the log-likelihood statistic are not reported because under all conditions the correct number of latent classes was not found. 

The difference in correct enumeration selection is obvious between the analyses for all information criteria considered. By assuming the indicators are continuous, the four information criteria are typically unable to find the correct number of latent classes under these conditions. The only exception is the BIC under the control condition when the response distribution is fairly normally distributed; yet, the identification rate is still low (see Table \ref{tb:results}). When the assumption of categorical indicators is used, the rate of correct specification increases remarkably. For example, the ssBIC perfectly identified the number of latent classes in all conditions under the assumption of categorical indicators, but under the assumption of continuity, the ssBIC performed poorly (see Table \ref{tb:results}). These results lend evidence to how sensitive the information criteria are to the assumption of continuity. 
\begin{table}[ht]
\centering
\footnotesize
\begin{threeparttable}
\caption{Proportion of data sets each fit measure correctly identified a three class solution as optimal by analysis type}
\label{tb:results}
\begin{tabular}{ lll *{12}{c}}
  \toprule
   &	&	&	 \multicolumn{2}{c}{AIC} &	\multicolumn{2}{c}{AICc}  &	\multicolumn{2}{c}{BIC} &	\multicolumn{2}{c}{ssBIC} &	\multicolumn{2}{c}{aLMR}  &	\multicolumn{2}{c}{BLRT} \\  \cmidrule(lr){4-5} \cmidrule(lr){6-7} \cmidrule(lr){8-9} \cmidrule(lr){10-11} \cmidrule(lr){12-13}  \cmidrule(lr){14-15}
RF\tnote{1} & CP\tnote{2} & N  & LCA & LPA & LCA & LPA & LCA & LPA & LCA & LPA  & LCA & LPA  & LCA & LPA \\
  \midrule
U & 1 & 500 & .75 & .00 & 1.00 & .00 & 1.00 & .00 & 1.00 & .00 & .58 & {\bf .69} & .69 & .00 \\ 
   &  & 1000 & .74 & .00 & 1.00 & .00 & 1.00 & .00 & 1.00 & .00 & .43 & .27 & .61 & .00 \\ 
   &  & 1500 & .72 & .00 & .98 & .00 & 1.00 & .00 & 1.00 & .00 & .35 & .12 & .61 & .00 \\ 
   & 2 & 500 & .72 & .00 & 1.00 & .00 & 1.00 & .00 & 1.00 & .00 & .70 & .64 & .69 & .00 \\ 
   &  & 1000 & .70 & .00 & .99 & .00 & 1.00 & .00 & 1.00 & .00 & .47 & .15 & .62 & .00 \\ 
   &  & 1500 & .67 & .00 & .96 & .00 & 1.00 & .00 & 1.00 & .00 & .37 & .03 & .58 & .00 \\ 
   & 3 & 500 & .75 & .00 & 1.00 & .00 & .98 & .00 & 1.00 & .00 & .66 & .45 & .83 & .00 \\ 
   &  & 1000 & .70 & .00 & .99 & .00 & 1.00 & .00 & 1.00 & .00 & .49 & .07 & .76 & .00 \\ 
   &  & 1500 & .72 & .00 & .97 & .00 & 1.00 & .00 & 1.00 & .00 & .34 & .01 & .73 & .00 \\ 
    ~~\\
  I & 1 & 500 & .86 & .00 & 1.00 & .00 & 1.00 & .00 & 1.00 & .00 & .01 & {\bf .69} & .67 & .00 \\ 
   &  & 1000 & .83 & .00 & 1.00 & .00 & 1.00 & .00 & 1.00 & .00 & .25 & {\bf .31} & .69 & .00 \\ 
   &  & 1500 & .81 & .00 & 1.00 & .00 & 1.00 & .00 & 1.00 & .00 & .56 & .12 & .70 & .00 \\ 
   & 2 & 500 & .90 & .00 & 1.00 & .00 & 1.00 & .00 & 1.00 & .00 & .07 & {\bf .66} & .67 & .00 \\ 
   &  & 1000 & .82 & .00 & 1.00 & .00 & 1.00 & .00 & 1.00 & .00 & .48 & .20 & .66 & .00 \\ 
   &  & 1500 & .76 & .00 & .99 & .00 & 1.00 & .00 & 1.00 & .00 & .77 & .04 & .60 & .00 \\ 
   & 3 & 500 & .89 & .00 & 1.00 & .00 & .96 & .00 & 1.00 & .00 & .09 & {\bf .46} & .78 & .00 \\ 
   &  & 1000 & .83 & .00 & 1.00 & .00 & 1.00 & .00 & 1.00 & .00 & .25 & .13 & .79 & .00 \\ 
   &  & 1500 & .80 & .00 & 1.00 & .00 & 1.00 & .00 & 1.00 & .00 & .38 & .02 & .76 & .00 \\ 
  ~~\\
  C & 1 & 500 & .63 & .00 & 1.00 & .00 & 1.00 & .45 & 1.00 & .00 & .58 & {\bf .71} & .87 & .00 \\ 
   &  & 1000 & .65 & .00 & 1.00 & .00 & 1.00 & .04 & 1.00 & .00 & .78 & .43 & .83 & .00 \\ 
   &  & 1500 & .66 & .00 & .99 & .00 & 1.00 & .00 & 1.00 & .00 & .92 & .25 & .80 & .00 \\ 
   & 2 & 500 & .64 & .00 & 1.00 & .00 & 1.00 & .14 & 1.00 & .00 & .72 & .67 & .84 & .00 \\ 
   &  & 1000 & .62 & .00 & 1.00 & .00 & 1.00 & .00 & 1.00 & .00 & .89 & .21 & .81 & .00 \\ 
   &  & 1500 & .62 & .00 & .99 & .00 & 1.00 & .00 & 1.00 & .00 & .96 & .07 & .78 & .00 \\ 
   & 3 & 500 & .65 & .00 & 1.00 & .00 & .88 & .11 & 1.00 & .00 & .43 & {\bf .57} & .91 & .00 \\ 
   &  & 1000 & .64 & .00 & 1.00 & .00 & 1.00 & .00 & 1.00 & .00 & .48 & .16 & .90 & .00 \\ 
   &  & 1500 & .66 & .00 & .99 & .00 & 1.00 & .00 & 1.00 & .00 & .57 & .03 & .86 & .00 \\ 
   \bottomrule
\end{tabular}
	\begin{tablenotes}[para, flushleft]
    {\small
        \textit{Note.}Cells in boldface indicate that LPA correctly specified the correct number of class more often than LCA. LCA = latent class analysis; LPA = latent profile analysis; AIC = Akaike information criterion; AICc = corrected Akaike information criterion; BIC = Bayesian information criterion; ssBIC = sample size-adjusted Bayesian information criterion; aLMR = adjusted Lo - Mendell - Rubin likelihood ratio test; BLRT = bootstrap likelihood ratio test. \\
	\item[1] RF is the response frequency condition described in Table \ref{tb:respdist}, where U = unendorsed category, I = infrequently endorsed category, and C = control condition;\\
	 \item[2] CP is class prevalence condition, 1) .45, .40, .15; 2) .59, .26, .15; 3) .89, .08, .03.
	} 
	\end{tablenotes}
	\end{threeparttable}
\end{table}

\subsection*{Likelihood Ratio Tests}
Two likelihood ratio tests for model selection were examined, the adjusted Lo-Mendell-Rubin likelihood ratio test (aLMR) and the bootstrap likelihood ratio test (BLRT). Both likelihood ratio tests performed worse than the information criteria included. The assumption of continuity had the most effect on the performance of the BLRT, where the BLRT did not identify the correct number of latent classes under all conditions (see Table \ref{tb:results}). Although, the BLRT performed consistently when the indicators are assumed to be categorical. The stark difference of BLRT performance between LCA and LPA is not found with the aLMR.

The performance of aLMR is more nuanced in the studied conditions compared to the other fit measures. We examined the performance of the aLMR with logistic regression to help identify which factors and levels included in our study are influential in determining correct specification. Notable results are evidence of an interaction between analysis type-LPA and response distribution-infrequently endorsed category (OR = 124.64); and an interaction among the analysis type (LPA), response distribution-infrequently endorsed category, class prevalence (.89 - .08 - .03) and sample size-1000 (OR = 8.17). 

\begin{figure}
\includegraphics[width = \textwidth]{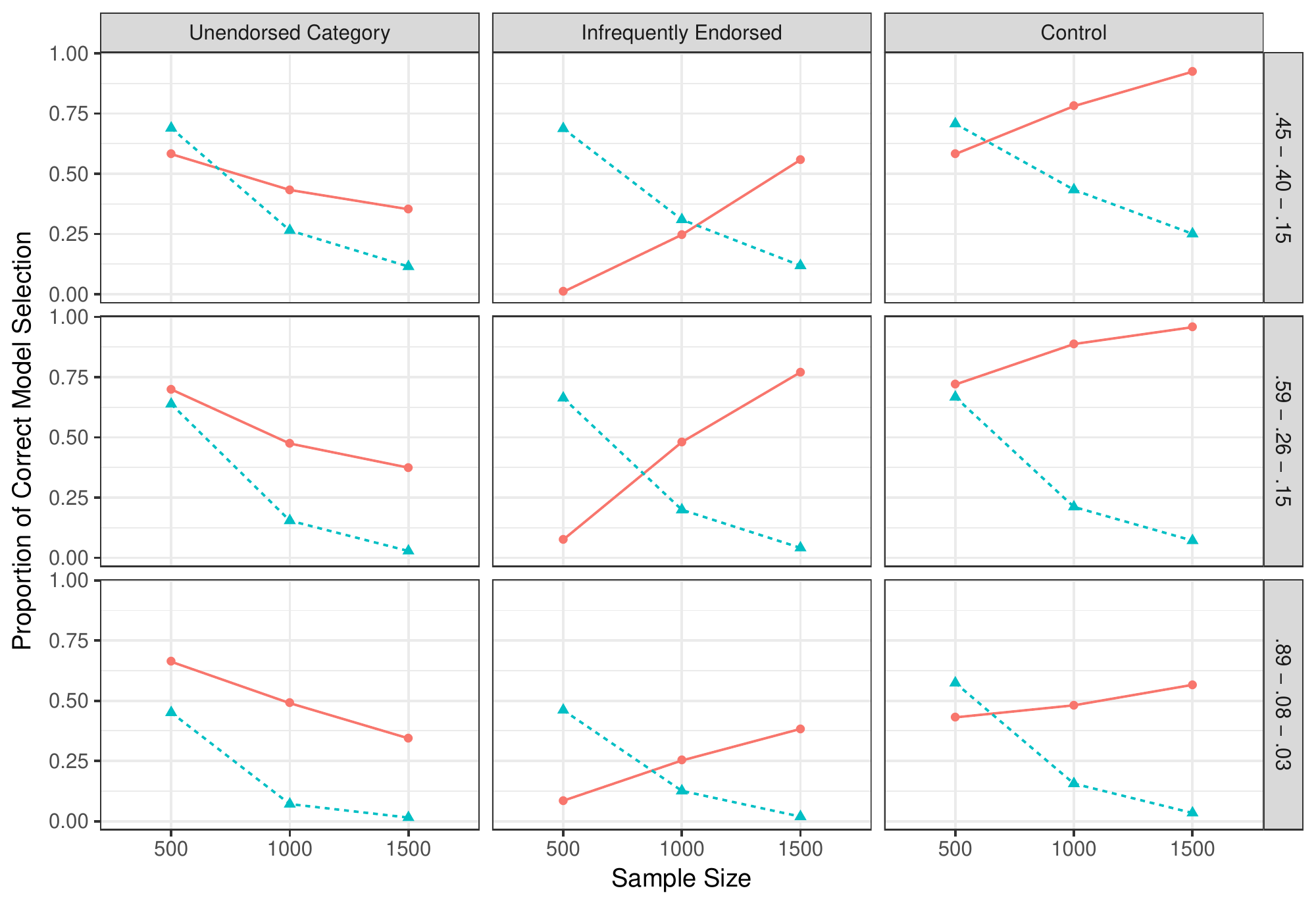}
\caption{Complex interactions of the aLMR results for estimating the correct number of latent classes}
\label{fig:almr}
\textit{Note.} The solid line represents LCA while the dashed line represents LPA results. The vertical panels are for each condition of class prevalence and the horizontal panels are for the response frequency conditions.
\end{figure}

The complex interactions lend evidence that the aLMR performs better under some conditions and when different assumptions of the continuity of the indicators are made. These interaction are difficult to make sense of from looking at the values reported in Table \ref{tb:results}, so the proportion of replications where the aLMR identified a three class solution as optimal is plotted by each condition and analysis type in Figure \ref{fig:almr} to aid interpretation. The assumption of continuity does appear to help when a response category is infrequently endorsed and sample size is small, but as sample size increases the assumption of continuity does not appear to help with model selection. 

\section*{Discussion}

The use of classification analyses is multidisciplinary so determining proper uses of available techniques and how to evaluate obtained results is crucial. 
The aim of our investigation was to find evidence of what occurs in model selection when ordered response categories are assumed to be continuous.
We studied what statistical measures of model fit are useful when data are ordered response categories and when a response category is unendorsed or infrequently endorsed under the different assumptions of continuity of indicators. 

The use of ordered response categories can lead to possible extraction issues. When ordered categories are assumed to be continuous, extraction of an incorrect number of latent classes is more likely. 
These ordered categorical data are highly prevalent in applied research in education and psychology, as found in a review by \cite{Harwell2001}. An accepted guideline in methodological research is that these data can be treated as continuous if there are at least five categories. 
However, in conditions studied here for mixture modeling we found evidence that five ordered response categories should be treated as ordered categorical and not continuous. 
The assumption of continuity only seems beneficial when convergence is problematic, such as when a response category is infrequently endorsed.  
When convergence is not an issue, our results are similar to previous simulation studies of these fit indices. The BIC and ssBIC were found to consistently select the correct number of latent classes, a result similar to other studies \citep{Morgan2015, Nylund2007, Burnham2004, Soromenho1994}. The BLRT was also found to perform poorly, which is counter to what \cite{Nylund2007} found but is similar to the results of \cite{Dziak2014}. 

These statistical measures of model selection appears to be sensitive to the assumption of continuity and the frequency of response category endorsement. This implies that there are no strict rules or guidelines for use of these fit indices, and researchers should be aware of possible issues of relying on these indices too heavily. Despite the fact that these measures can be useful, performance of these fit indices changes based on the distribution pf the indicators and the assumptions of the researcher.

\subsection*{Limitations and Delimitations}

As with any simulation study, these results generalize only to the conditions included in this study. We used default settings in M{\it plus} 8.1 to try to generalize as widely as possible. However, additional settings in M{\it plus} could help with issues that arise with estimation. For example, Bayesian estimation procedures are a promising avenue \citep{Asparouhov2017}. The included statistical fit indices are only a subset of possible choices available to practitioners; therefore the generalizations of our investigation for model selection is limited to these included measures. 

\subsection*{Recommendations for Use of Fit Indices in LCA \& LPA}
We recommend that when data collected are five ordered categories, such as Likert-type data, and are being used for identification of homogeneous subpopulations, these data should be considered categorical and not continuous. Despite the advantage of convergence, the commonly available fit statistics for model selection performed poorly under the assumption of continuity with these data. One exception is the adjusted Lo-Mendell-Rubin likelihood ratio statistic. The aLMR is the only fit measure examined in this study to identify the correct number of latent classes when a response category is infrequently or not endorsed and the indicators are assumed to be continuous. The use of aLMR could help practitioners when model convergence is an issue and sample size is small. Although, under most conditions we recommend that the corrected AIC (AICc), Bayesian information criterion (BIC), or the sample size-adjusted BIC (ssBIC) be used to aid model selection when the indicators are five ordered categories along with substantial weight to substantive theory and interpretation of results.

\subsection*{Author Contributions}

R.N. Padgett and R. J. Tipton jointly generated the idea for the study. R.N. Padgett wrote the code for generating the data, estimating the models, extracting the relevant results and analyzing the model summaries. R. J. Tipton verified the accuracy of those analyses. R.N. Padgett and R. J. Tipton collaboratively wrote the first draft of the manuscript, and both authors critically edited it. Both authors approved the final submitted version of the manuscript.
% ====================== %
\newpage
\raggedright
%\bibliographystyle{apacite} 
% You may have to select another style. Remember: LaTeX, BibTeX, LaTeX, LaTex to get the citations to appear
%\raggedright
%\urlstyle{same}
%\bibliography{references}
\printbibliography
%% ====================== %
%
\end{document}